\documentclass[12pt]{article} 

\usepackage{ol2}
\usepackage[draft]{hyperref}
\usepackage{amsmath}
\usepackage[applemac]{inputenc}

\begin{document}

\title{Higher-order Kerr improve quantitative modeling of laser filamentation}

\author{M. Petrarca$^{1}$, Y. Petit$^{1}$, S. Henin$^{1}$, R. Delagrange$^{1}$, P. B\'ejot$^{1,2}$, J. Kasparian$^{1,*}$}

\address{(1) GAP-Biophotonics, Universit\'e de Gen\`eve, Chemin de Pinchat 22, 1211 Geneva 4, Switzerland \\
(2) Laboratoire Interdisciplinaire Carnot de Bourgogne (ICB), UMR 5209 CNRS-Universit\'e de Bourgogne, BP 47870, F-21078 Dijon Cedex, France}
\email{* jerome.kasparian@unige.ch}

\begin{abstract}
We test numerical filamentation models against experimental data about the peak intensity and filament density in laser filaments. We show that the consideration of the higher-order Kerr effect (HOKE) improves the quantitative agreement without the need of adjustable parameters.
\end{abstract}



\noindent Laser filamentation \cite{BraunKLDSM1995,CouaironM07,Chin,Kasparian} is a propagation regime typical of high-power lasers. It stems from a dynamic balance between self-focusing and self-defocusing non-linearities of different orders, i.e., different intensity dependences. Self-focusing by the Kerr effect is well established. Conversely, the relative contributions to defocusing are still discussed. The two main effects nowadays considered in gases are the laser-generated plasma and the saturation of the medium polarisability under strong-field illumination \cite{PopovJETP,IvanovPNAS,PierreTDSE}, empirically described as a power series of the intensity and referred to as higher-order Kerr effect (HOKE).
The latter have been first introduced as freely adjustable parameters \cite{Hovha2001,VincoB2004}, before experimental values were reported \cite{Loriot09,arXivLoriot}. 

Strong effort have been dedicated to reach a quantitative agreement between numerical models of non-linear pulse propagation and experimental data about filamentation \cite{SkupiSBLSSZS2006,ChampBGTPS2008,KolesMDM2010,PolynKWM2011}. In that purpose, a common approach consists in tweaking simulation inputs like the non-linear index or the ionization rates, and/or the experimental parameters such as the beam energy, pulse duration, beam diameter or profile, until a satisfactory match with the experiments is achieved.
The remaining discrepancies are generally attributed, among others, to the difficulty of the experimental measurements inside the filaments, and the associated uncertainties. However, the introduction of new techniques, e.g. for measuring the peak intensity from the ratio of nitrogen fluorescence lines \cite{DaiglJHWKRBC2010,XuSZCZLCXC2012} enhances the precision and renew the challenge for theoretical models to precisely match those data.

Here, we provide experimental electron density measurements over a wide range of pulse durations, and compare these experimental results with numerical simulations. We show that  considering  the HOKE in the model allows a good agreement between experimental results and simulations without any parameter adjustment. Such agreement is also confirmed when comparing the model including the HOKE \cite{Bejot} with previously published data about the filament peak intensity  \cite{XuSZCZLCXC2012} and electron density  \cite{ThebeLSBC2006}.

\begin{figure} [t!]
   \begin{center}
      \includegraphics[keepaspectratio,width=7.5cm]{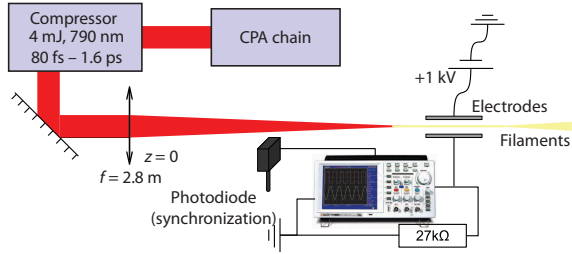}
   \end{center}
   \caption{Experimental setup}
   \label{setup}
\end{figure}

The filament ionization was characterized as a function of the pulse duration by using 4 mJ laser pulses centered at 800 nm, with 100 Hz repetition rate and 3 cm initial beam diameter. The slightly diverging beam was focused by an $f$=2.8 m lens to form a 20 cm long filament in air. The pulse duration was varied from 80 fs to 1.6 ps by detuning the grating compressor of the chirped pulse amplification (CPA) chain. The charge is measured by propagating the filaments between a pair of parallel electrodes (1 x 1 cm; 1 cm spacing) under 1 kV bias: the transient current through the circuit is proportional to the electron density generated by the filaments \cite{HeninPKKW2009a}, although with an arbitrary calibration factor (See Figure \ref{setup}). This charge measurement was double-checked by sonometric measurements \cite{YuMKSGFBW2003}. 

To get information about the role of the HOKE, we performed numerical simulations considering a linearly polarized incident electric field at a wavelength $\lambda_0$=$800$~nm with cylindrical symmetry around the propagation axis $z$. According to the unidirectional propagation pulse equation (UPPE) \cite{Moloney_UPPE}, the scalar envelope $\varepsilon(r,t,z)$ (defined such that $|\varepsilon(r,z,t)|^2=I(r,z,t)$, $I$ being the intensity) evolves in the frame traveling at the pulse velocity according to \cite{B'ejHLKWF2011a}:
\begin{eqnarray}
\begin{aligned}
\partial_z\widetilde{\varepsilon}=&i(\sqrt{k^2(\omega)-k_\bot^2}-k'\omega)\widetilde{\varepsilon}&\\
&+\frac{1}{\sqrt{k^2(\omega)-k_\bot^2}}\left(\frac{i\omega^2}{c^2}\widetilde{P}_{\textrm{NL}}-\frac{\omega}{2\epsilon_0c^2}\widetilde{J}\right)-\widetilde{\alpha},&
\label{Eq.final_enveloppe}
\end{aligned}
\end{eqnarray}
where $c$ is the velocity of light in vacuum, $\omega$ is the angular frequency, $k(\omega)$=n($\omega$)$\omega$/c, $k'$ its derivative at $\omega_0$=2$\pi$c/$\lambda_0$, $n(\omega)$ is the linear refractive index at the frequency $\omega$, $k_{\bot}$ is the spatial angular frequency. $P_{\textrm{NL}}$ is the nonlinear polarization, $J$ is the free-charge induced current and $\alpha$ is the nonlinear losses induced by photo-ionization. $\widetilde{f}$ denotes simultaneous temporal Fourier and spatial Hankel transforms of function $f$.
The non-linear polarization is evaluated as $P_{\textrm{NL}}=\left(\sum_j{n_{2j}|\varepsilon|^{2j}}+\Delta n_{\textrm{Raman}}\right)\varepsilon$, where the $n_{2j}$ are the $j^{\textrm{th}}$-order nonlinear refractive indices ($n_2$=1.2$\times$10$^{-7}$ cm$^2$/TW, $n_4$=-1.5$\times$10$^{-9}$ cm$^4$/TW$^2$,
 $n_6$=2.1$\times$10$^{-10}$ cm$^6$/TW$^3$, 
$n_8$=-8$\times$10$^{-12}$ cm$^8$/TW$^4$ \cite{Loriot09}) and $\Delta n_{\textrm{Raman}}$ is the Raman induced refractive index change evaluated by solving the rotational time-dependent Schr\"odinger equation of both N$_2$ and 0$_2$ in the weak field regime. Alternatively, the model is truncated to the third-order non-linearity (i.e. the $n_2 |\varepsilon|^2\varepsilon$ term) to disregard the contribution of the HOKE.
The current is evaluated as $\widetilde{J}=\frac{e^2}{m_\textrm{e}}(\nu_\textrm{e}+i\omega)\widetilde{\rho\varepsilon}/(\nu_\textrm{e}^2+\omega^2)$, where $e$ and $m_\textrm{e}$ are the electron charge and mass respectively, $\epsilon_0$ is the vacuum permittivity, $\nu_\textrm{e}$ is the effective electronic collisional frequency, and $\rho$ is the {electron} density. Finally, $\alpha=\sum_{i=O_2,N_2}{W_i(|\varepsilon|^2)U_i\rho_{\textrm{at},i}/(2|\varepsilon|^2)}$, $\rho_{\textrm{at},i}$ is the density of molecules of species $i$, $W_i(|\varepsilon|^2)$ is their photoionization probability modeled by the PPT formulation, with ionization potential $U_i$.

The propagation dynamics of the electric field is coupled with the electron density $\rho$, calculated as \cite{CouaironM07}
\begin{equation}
\partial_t\rho=\sum_{i=O_2,N_2}{\left(W_i(|\varepsilon|^2)\rho_{at,i}+\frac{\sigma_i}{U_i}\rho|\varepsilon|^2\right)-\beta\rho^2},
\label{Eq.electron_dens}
\end{equation}
where $\beta$ are the electron recombination rate and $\sigma_i$ are the inverse Bremsstrahlung cross-sections with species $i$, also accounting for avalanche ionization. 
Note that our model relies only on published values \cite{Loriot09,Kasparian} and contains no adjustable parameter.
For each experimental situation, we directly used the experimental parameters, without any tweaking. Unless otherwise specified, Gaussian pulse shapes have been considered, both spatially and temporally.

\begin{figure} [tb]
   \begin{center}
      \includegraphics[keepaspectratio,width=7.5cm]{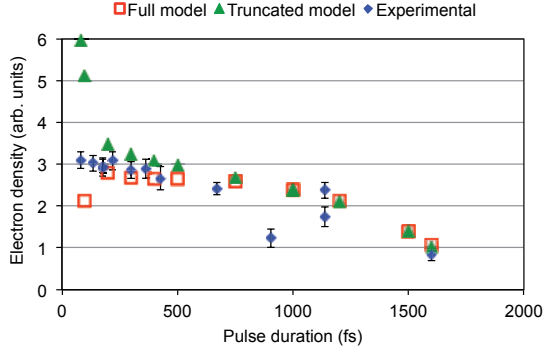}
   \end{center}
   \caption{Effect of pulse duration on the electron density in filaments. Each numerical curve is normalized independently to optimally match the experimental results.}
   \label{Raphaelle}
\end{figure}

The measured electron density rose very slowly (less than a factor of 4) when the pulse duration was decreased from 1.6 ps to 80 fs, although the peak power drastically increased by a factor of 20 (Figure \ref{Raphaelle}). This might seem unexpected when considering that the ionization rate scales with $I^8$ in oxygen \cite{TalebYC1999,Kasparian2}. This relative stability can however be understood by considering that the free electrons need several picoseconds to tens of ps to recombine, so that they accumulate more efficiently during longer pulses, an effect that partly balances the lower ionization rates. While both models reproduce this slow variation for pulses below a few hundreds of femtoseconds, the full model seems slightly more accurate for shorter pulses.


\begin{figure} [tb]
   \begin{center}
      \includegraphics[keepaspectratio,width=8.6cm]{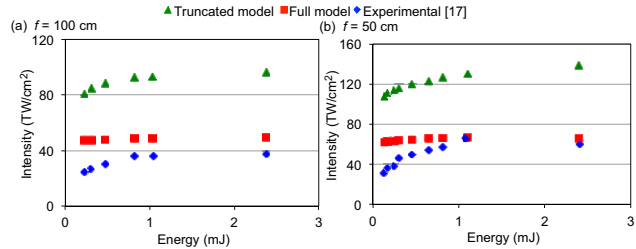}
   \end{center}
   \caption{Experimental \cite{XuSZCZLCXC2012} and simulated peak intensity in filaments generated by 42 fs pulses of 1 cm diameter, focused with (a) $f$ = 1 cm and (b) $f$=50 cm.}
   \label{Intensite_Chin_Opex}
\end{figure}

To get a more direct comparison between the two models, we successively tested them against previously published \emph{quantitative} experimental data about the peak intensity  \cite{XuSZCZLCXC2012} and electron density \cite{ThebeLSBC2006} measured in filaments by different groups and techniques, so as to avoid individual artifacts or systematic errors to affect our conclusions.

As shown in Figure \ref{Intensite_Chin_Opex}, the peak intensity of filaments generated by focused 42 fs pulses of 1 cm diameter at 1/$e^2$ (i.e, 5.9 mm FWHM) measured via the ratio of nitrogen fluorescence lines \cite{XuSZCZLCXC2012} is much better reproduced by the full model than by the truncated one. The latter, in which the HOKE defocusing terms are disregarded, overestimates the intensity by a factor of typically 2--3.

\begin{figure} [tb]
   \begin{center}
      \includegraphics[keepaspectratio,width=8.6cm]{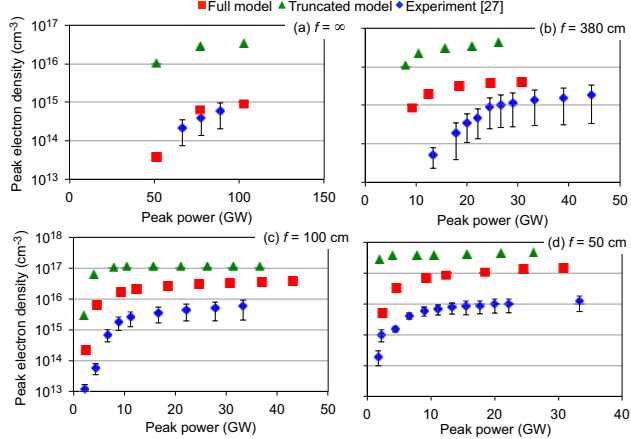}
   \end{center}
   \caption{Experimental \cite{ThebeLSBC2006} and simulated electron density as a function of focusing and incident power, for a 45 fs pulse of initial beam diameter of 8.4 mm (full width at 1/$e^2$), for (a) a free-propagating (parallel) beam; (b) $f$ = 380 cm; 
   (c)  $f$ = 100 cm; 
   (d) $f$ = 50 cm.
   }
\label{Chin_electrons}
\end{figure}

The electron density as measured in \cite{ThebeLSBC2006} provides a much more sensitive variable, because its high-order dependence on intensity magnifies the experiment dynamics, allowing more reliable comparisons.
The intensity of the nitrogen fluorescence constitutes a classical measurement of the electron density \cite{HosseYLC2004}. Consistent with the differences in the predicted intensities, the full model always yields a 4- to 30-times lower electron density  than the truncated one.
The match with experimental data \cite{ThebeLSBC2006} is perfect for a parallel beam corresponding to typical filamentation conditions (Figure \ref{Chin_electrons}a). However (Figure \ref{Chin_electrons}b--d), even the full model tends to overestimate the electron density produced by focused beams. This discrepancy could be due to the recently suggested overestimation \cite{PierreTDSE} of the free electron density by the PPT model \cite{TalebYC1999,Kasparian}, or to optical aberrations and astigmatism in experiments, which are not included in the model and tend to decrease the peak intensity in the waist region \cite{SunJLZYG2005} and reduce filament length and strength\cite{KamaliSDABC2008}. 
Besides, this increasing geometrical constraint explains the decrease of the difference between the two models for tighter focusing. Therefore, the most concluding conditions to test filamentation models correspond the less focused ones, i.e. the larger $f$ number.


Our results show that considering the saturation of the medium polarization in the strong-field regime (aka HOKE) improves the quantitative modeling of filamentation, especially for collimated beams for which the geometrical constraint does not interfere with the self-guiding process. Such update of the standard laser filamentation model is easy to implement, based on both experimental measurement \cite{Loriot09} and quantum mechanical justifications \cite{PopovJETP,IvanovPNAS,PierreTDSE}. It  will in particular impact the determination of the optimal conditions for the potential atmospheric applications \cite{Kasparian_Science,Kasparian_Opex} of laser filamentation, like rainmaking \cite{Rain_Nature}, or lightning control \cite{Lightning_Opex}.

\textbf{Acknowlegments.}
This work was supported by the European Research Council Advanced Grant "Filatmo" and the Conseil R\'egional de Bourgogne (FABER program).
We thank F. Th\'eberge for their assistance in accessing their experimental data reproduced in Figure \ref{Chin_electrons} \cite{ThebeLSBC2006}.

\end{document}